\documentclass[aps,prl,twocolumn,superscriptaddress,preprintnumbers,floatfix]{revtex4}
\usepackage{graphics,colordvi}
\usepackage{amsmath,amssymb,epsfig}

\makeatletter
\newcommand{\fmslash}[2][0mu]{%
  \mathchoice
    {\fmsl@sh\displaystyle{#1}{#2}}%
    {\fmsl@sh\textstyle{#1}{#2}}%
    {\fmsl@sh\scriptstyle{#1}{#2}}%
    {\fmsl@sh\scriptscriptstyle{#1}{#2}}}
\newcommand{\fmsl@sh}[3]{%
  \m@th\ooalign{$\hfil#1\mkern#2/\hfil$\crcr$#1#3$}}

\newcommand{\tr}{\hbox{tr}}

\def\eq#1\en{\begin{equation} #1 \end{equation}}

\def\eqa#1\ena{\begin{eqnarray} #1 \end{eqnarray}}

\newcommand{\GeV}{\mbox{\rm GeV}}

\usepackage{graphicx,color}

\begin{document}

\title{Photon-neutrino interaction in $\theta$-exact covariant noncommutative model:\\
Phenomenology and Quantum properties*}
\author{Josip Trampeti\'{c}}
\affiliation{Rudjer Bo\v skovi\' c Institute, P.O.Box 180, HR-10.002 Zagreb, Croatia}
\affiliation{Max-Planck-Institut f\"ur Physik,
	(Werner-Heisenberg-Institut),
  	 F\"ohringer Ring 6, D-80805 M\"unchen, Germany}


\begin{abstract}
We construct the $\theta$-exact covariant noncommutative (NC) 
model and obtain various closed constraints on 
the NC scale ($\Lambda_{\rm NC}$) from inelastic neutrino-nucleon scatterings ($\nu N \to \nu+X$), 
from plasmon decay into neutrino pair ($\gamma_{pl}\to\bar\nu\nu$), 
from the BBN nucleosynthesis and from the reheating phase after inflation, respectively. 
We have have found neutrino two-point function in a closed form.  It shows decoupling of 
the UV divergent term from softened UV/IR mixing term and from the finite terms as well. 
We show that problematic UV divergent and UV/IR mixing terms vanish
for a certain choice of the noncommutative parameter $\theta$ which preserves unitarity.
We find non-perturbative modifications of the neutrino 
dispersion relations which are assymptotically independent
of the scale of noncommutativity in both the low and high energy 
limits and may allow superluminal propagation.
\end{abstract}

\maketitle

{\it Introduction}\\
The  study of spacetime quantization has  originally been motivated by  
major problems of physics at
extremely-high energies, in particular the problems of renormalization 
and quantum gravity. Heisenberg-type spacetime uncertainty relations 
can effectively lead to a replacement of the
continuum of points by finite size spacetime cells, thus providing a means 
by which to tame UV divergences.
It is reasonable to expect that noncommutative~(NC) field theory models can
provide  some guidance for a deeper understanding of the structure of spacetime  
at extremely-high energies. In fact, these NC models appear quite naturally 
in string theory~\cite{Seiberg:1999vs}.  
The relevant NC scale may very well be beyond direct experimental 
reach for the foreseeable future 
(except in certain theories with large extra dimensions). 
Nevertheless, non-perturbative effects can nevertheless lead 
to profound observable consequences for low energy physics. 
A famous example is UV/IR mixing \cite{Minwalla:1999px,Matusis:2000jf}. 
Another striking example is 
the running of the coupling constant of NC QED \cite{MS-R}, 
rendering that theory asymptotically free.
In this article we derive similarly non-perturbative result 
for $\nu$-propagation in the NC spacetimes, 
including in particular modified dispersion relations 
that allow superluminal velocities as they are currently discussed 
in the context of the OPERA experiment~\cite{:2011zb}.

In a simple model of NC spacetime local coordinates
$x^\mu$ are promoted to hermitian operators
$\hat x^\mu$ satisfying spacetime NC and implying 
uncertainty relations
\begin{equation}
[\hat x^\mu ,\hat x^\nu]=i\theta^{\mu\nu}
\longrightarrow \Delta x^\mu \Delta x^\nu \geq \frac{1}{2}|\theta^{\mu\nu}|,
\label{x*x}
\end{equation}
where $\theta^{\mu\nu}$ is real, antisymmetric matrix. 

The Moyal-Weyl $\star$-product, relevant for the case of 
a constant $\theta^{\mu\nu}$, is defined as follows:
\begin{equation}
(f\star g)(x)=f(x)
e^{\frac{i}{2}\overleftarrow{{\partial}_\mu}\,
\theta^{\mu\nu}\,\overrightarrow{{\partial}_\nu}} g(x).
\label{f*g}
\end{equation}
The operator commutation relation (\ref{x*x}) is then realized by the $\star$-commutator
\begin{equation}
[\hat x^\mu ,\hat x^\nu]=[x^\mu \stackrel{\star}{,} x^\nu]=i\theta^{\mu\nu}.
\end{equation}
In our construction the noncommutative fields are 
obtained via so-called Seiberg-Witten (SW) maps~\cite{Seiberg:1999vs}  
from the original commutative fields.
It is commutative instead of the noncommutative gauge symmetry that is
preserved as the fundamental symmetry of the theory.

The perturbative quantization of noncommutative field theories was first
proposed in a pioneering paper by Filk \cite{Filk:1996dm}.
However, despite of some significant progress like the
models in \cite{Grosse:2004yu} and \cite{arXiv:1111.5553}, a better understanding
of various models quantum loop corrections still remains in general a challenging open question. This fact is particularly true for the models constructed
by using SW  map expansion in the NC parameter $\theta$, including NCSM
\cite{Madore:2000en,Calmet:2001na,Aschieri:2002mc,Schupp:2002up}. 
However the resulting models are very useful as effective field theories
including relevant phenomenology 
\cite{Melic:2005hb,Ohl:2004tn,Ohl:2010zf,Buric:2007qx}
and their one-loop quantum properties \cite{Bichl:2001cq,Martin:2002nr,Brandt:2003fx,Buric:2006wm,Latas:2007eu,Martin:2007wv,Buric:2007ix,Martin:2009sg,Martin:2009vg,Tamarit:2009iy,Buric:2010wd}.
 
Quite recently, $\theta$-exact SW map expansions,
in the framework of covariant noncommutative quantum gauge field theory \cite{Jackiw:2001jb}, were applied in loop computation 
\cite{Zeiner:2007,Schupp:2008fs,arXiv:1109.2485,arXiv:1111.4951}
and phenomenology \cite{Horvat:2010sr,Horvat:2011iv}.
\\

\begin{figure}[top]
\begin{center}
\includegraphics[width=8.5cm,height=5cm]{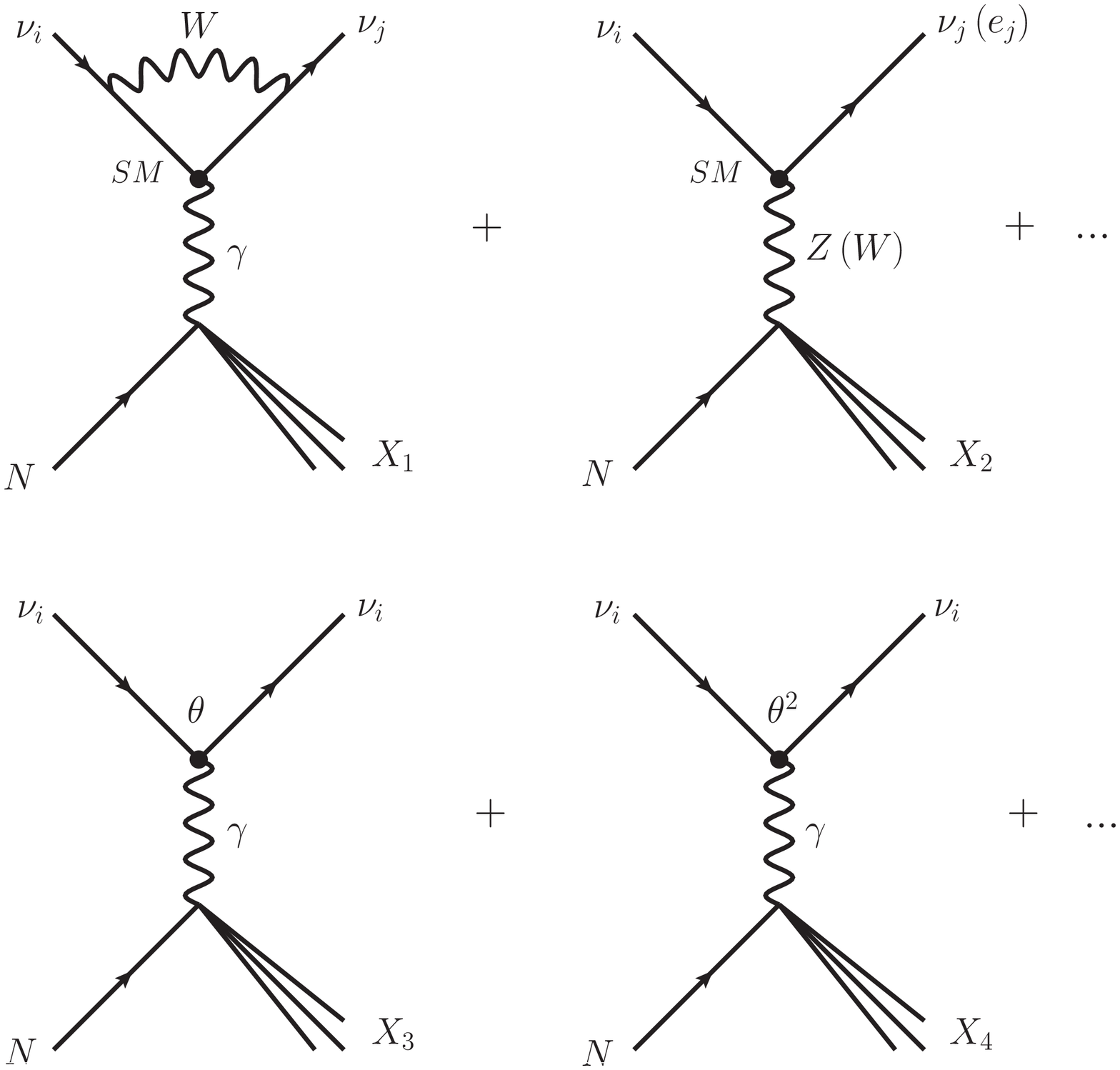}
\end{center}
\caption{Diagrams contributing to $\nu N\to\nu+X$ processes. }
\label{fig:trampslika}
\end{figure}

{\it 1. Phenomenological motivation}\\
Our main motivation to discuss a constraint on the scale of the NC guge field 
theory (NCGFT), $\Lambda_{\rm NC}$, arising from 
ultra-high energy cosmic ray experiments involving 
$\nu$-nucleon inelastic cross section,
is physically due to the possibility of a direct coupling of 
$\nu$'s to gauge bosons in the NC background \cite{Schupp:2002up}, which 
plays the role of an external field in the theory  (see i.e. Fig. \ref{fig:trampslika}).
\begin{figure}[top]
\begin{center}
\includegraphics[width=8.5cm,height=5cm]{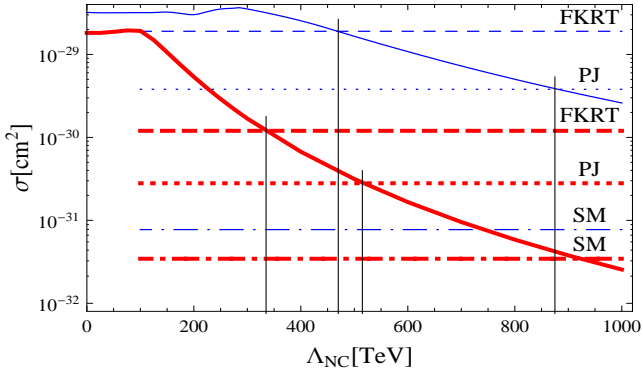}
\end{center}
\caption{$\nu N\to\nu\,+\, anything$ cross sections
vs. $\Lambda_{\rm NC}$ for $E_\nu=10^{10}~\GeV$ (thick lines) 
and $E_\nu=10^{11}~\GeV$ (thin lines). FKRT and PJ lines are the upper
bounds on the $\nu$-nucleon inelastic cross section, denoting different 
estimates for the cosmogenic $\nu$-flux. SM denotes the SM total
(charged current plus neutral current) $\nu$-nucleon inelastic cross
section. The vertical lines denote the intersections of our curves with the
RICE results.  }
\label{fig:ncSM-CrossSections}
\end{figure}
\begin{figure}[top]
\begin{center}
\includegraphics[width=8cm,height=5cm]{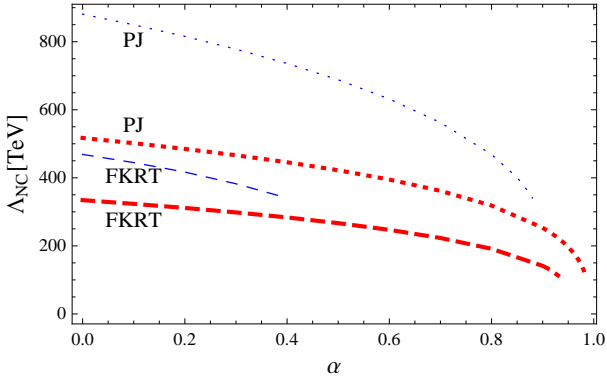}
\end{center}
\caption{The intersections of our curves with the RICE results (cf. Fig.2)
as a function of the fraction of Fe nuclei in the UHE cosmic rays. The
terminal point on each curve represents the highest fraction of Fe nuclei
above which no useful information on $\Lambda_{\rm NC}$ can be inferred with
our method.}
\label{fig:ncSM-LambdaVsAlpha}
\end{figure}
The observation of ultra-high energy (UHE) $\nu$'s from extraterrestrial
sources would open a new window to look to the cosmos, as such $\nu$'s may
easily escape very dense material backgrounds around local astrophysical
objects, giving thereby information on regions that are otherwise hidden to
any other means of exploration. In addition, $\nu$'s are not deflected on
their way to the earth by various magnetic fields, pointing thus back to
the direction of distant UHE cosmic-ray source candidates. This could also
help resolving the underlying acceleration in astrophysical sources.
    
Since the GZK-structure in the energy spectrum of UHE 
cosmic rays at $\sim 4 \times 10^{19}$ eV 
has been observed recently with high statistical accuracy \cite{Aglietta:2007yx}, 
the flux of the so-called 
cosmogenic $\nu$'s, arising from photo-pion production on the
cosmic microwave background $p \gamma_{CMB} \rightarrow \Delta^{*}
\rightarrow N\pi$ and subsequent pion decay, is now guaranteed to exist. 
Although estimates for the cosmogenic $\nu$-flux are very
model-dependent, primarily due to our insufficient knowledge of the nature
and the origin of UHE cosmic rays, possible ranges for the size of the
flux of cosmogenic $\nu$'s can be obtained from separate analysis of the data
from various large-scale observatories \cite{Fodor:2003ph,Protheroe:1995ft}. 

Considering the uncertainty in the flux of cosmogenic $\nu$'s regarding the
chemical composition of UHE cosmic rays see detailed 
discussion in \cite{Horvat:2010sr}. 
\begin{figure}
\begin{center}
\includegraphics[width=8cm,height=5.5cm]{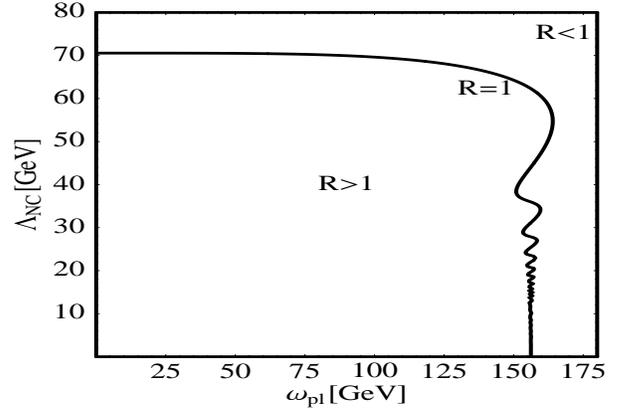}
\end{center}
\caption{The plot of scale $\Lambda_{\rm NC}$ versus the plasmon frequency 
$ \omega_{pl}$ with $R=1$}
\label{fig:RvsOmegaL}
\end{figure}
\begin{figure}
\begin{center}
\includegraphics[width=8cm,height=5.5cm]{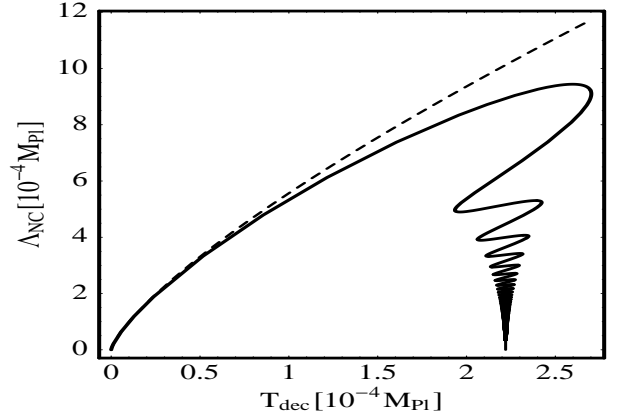}
\end{center}
\caption{The plot of the scale $\Lambda_{\rm NC}$ versus $T_{dec}$
for perturbative/exact solution ({dashed/full curve}).}
\label{fig:LambdaVsT}
\end{figure}
Employing the upper bound on the $\nu N$ cross section derived from the RICE
Collaboration search results \cite{Kravchenko:2002mm} at $E_{\nu} =
10^{11}$ GeV ($4\times 10^{-3}$ mb for the FKRT $\nu$-flux \cite{Fodor:2003ph})), 
one can infer from $\theta$-truncated model    
on the NC scale $\Lambda_{\rm NC}$ to be greater than 455 TeV, 
a really strong bound. 
Here we have used $\theta^{\mu\nu} \equiv c^{\mu\nu}/\Lambda_{\rm NC}^2 $ 
such that the matrix elements of $c$ are of order one.

One should however be
careful and suspect this result as it has been obtained from the conjecture
that the $\theta$-expansion stays well-defined in the kinematical region of interest. 
Although a heuristic criterion for the validity of the perturbative $\theta$-expansion,
$\sqrt{s}/\Lambda_{\rm NC} \lesssim\;1$, with $s = 2 E_{\nu}M_N$, would
underpin our result on $\Lambda_{\rm NC}$, a more thorough inspection on
the kinematics of the process does reveal a  more stronger energy
dependence  $E_{\nu}^{1/2} s^{1/4}/ \Lambda_{\rm NC} \lesssim 1$.
In spite of an additional phase-space suppression for small $x$'s in  
the $\theta^2$-contribution \cite{Ohl:2004tn} of
the cross section relative to the $\theta$-contribution, we find an unacceptably 
large ratio $\sigma({\theta^2})/\sigma({\theta}) \simeq 10^4$,
at $\Lambda_{\rm NC}=455$ TeV.
Hence, the bound on $\Lambda_{\rm NC}$ obtained this way is incorrect, 
and our last resort 
is to modify the model adequately to include the full-$\theta$
resummation, thereby allowing us to compute nonperturbatively in $\theta$.

The behavior of the cross section
with the NC scale at fixed $E_{\nu} = 10^{10}$ GeV  and
$E_{\nu} = 10^{11}$ GeV, 
together with the upper bounds depending on the actual size of the
cosmogenic $\nu$-flux (FKRT \cite{Fodor:2003ph} and PJ \cite{Protheroe:1995ft}) 
as well as the total SM cross sections at these energies, are
depicted in our Figure \ref{fig:ncSM-CrossSections}. In order to maximize 
the NC $\theta$-exact effect we choose $c_{01}-c_{13}=c_{02}-c_{23}=c_{03}=1$.

If future data confirm that UHE cosmic rays are composed mainly of Fe
nuclei, as indicated, for the time being, by the PAO data, then still
valuable information on $\Lambda_{\rm NC}$ can be obtained with our method, as
seen in Fig.\ref{fig:ncSM-LambdaVsAlpha} . 
Here we see the intersections of our curves with the
RICE results (cf. Fig.\ref{fig:ncSM-CrossSections})
as a function of the fraction $\alpha$ of Fe nuclei in the UHE cosmic rays.
On top of results, presented in 
Figs.\ref{fig:ncSM-CrossSections} and \ref{fig:ncSM-LambdaVsAlpha},
we also have the NC scale given as a function of 
the plasmon frequency (Fig.\ref{fig:RvsOmegaL}) and as a function of 
the $T_{dec}$ from BBN (Fig.\ref{fig:LambdaVsT}), respectively. 
All results depicted in Figs.\ref{fig:ncSM-CrossSections}-\ref{fig:LambdaVsT},
shows closed/convergent behaviour.  In our opinion those were the strong signs   
to continue research towards  quantum properties of such $\theta$-exact 
noncommutative gauge field theory model. \\

{\it 2. Covariant $\theta$-exact ${\rm U_{\star}(1)}$ model }\\
Neutrinos do not carry an electromagnetic charge
and hence do not couple directly to photons. 
However, in the presence of spacetime NC, it is possible to couple photons 
to neutral  and ``chiral'' fermion particles \cite{Horvat:2010sr,Horvat:2011iv}.

We start with the following SW type of NC $\rm U_{\star}(1)$ gauge model:
\begin{equation}
S=\int-\frac{1}{4}F^{\mu\nu}\star F_{\mu\nu}+i\bar\Psi\star\fmslash{D}\Psi\,,
\label{S}
\end{equation}
with definitions of the nonabelian NC covariant derivative and
the field strength, respectively:
\begin{eqnarray}
D_\mu\Psi&=&\partial_\mu\Psi-i[A_\mu\stackrel{\star}{,}\Psi],
\nonumber\\
F_{\mu\nu}&=&\partial_\mu A_\nu-\partial_\nu
A_\mu-i[A_\mu\stackrel{\star}{,}A_\nu].
\label{DF}
\end{eqnarray}
All the fields in this action are images under (hybrid) Seiberg-Witten maps
of the corresponding commutative fields $a_\mu$ and $\psi$.
Here we shall interpret the NC fields as valued in
the enveloping algebra of the underlying gauge group.
This naturally corresponds to an expansion in powers of
the gauge field $a_\mu$ and hence in powers of
the coupling constant $e$. At each order in $a_\mu$ we shall
determine $\theta$-exact expressions.

In the next step we expand the action in terms of
the commutative gauge parameter $\lambda$
and fields $a_\mu$ and $\psi$ using the SW map solution \cite{Schupp:2008fs}
up to the $\mathcal O(a^3)$ order:
\begin{eqnarray}
A_\mu&=&\,a_\mu-\frac{1}{2}\theta^{\nu\rho}{a_\nu}\star_2(\partial_\rho
a_\mu+f_{\rho\mu}),
\nonumber\\
\Psi&=&\psi-\theta^{\mu\nu}
{a_\mu}\star_2{\partial_\nu}\psi
\nonumber\\
&+&\frac{1}{2}\theta^{\mu\nu}\theta^{\rho\sigma}
\bigg\{(a_\rho\star_2(\partial_\sigma
a_\mu+f_{\sigma\mu}))\star_2{\partial_\nu}\psi
\nonumber\\
&+&2a_\mu{\star_2}
(\partial_\nu(a_\rho{\star_2}\partial_\sigma\psi))
-a_\mu{\star_2}(\partial_\rho
a_\nu{\star_2}\partial_\sigma\psi)
\nonumber\\
&-&\big(a_\rho\partial_\mu\psi(\partial_\nu
a_\sigma+f_{\nu\sigma})
-\partial_\rho\partial_\mu\psi a_\nu
a_\sigma\big)_{\star_3}\bigg\},
\nonumber\\
\Lambda&=&\lambda-\frac{1}{2}\theta^{ij}a_i\star_2\partial_j\lambda \,,
\label{SWmap}
\end{eqnarray}
with $\Lambda$ being the NC gauge parameter and
$f_{\mu\nu}$ is the abelian commutative field strength
$f_{\mu\nu}=\partial_\mu a_\nu-\partial_\nu a_\mu$.

The generalized Mojal-Weyl star products
$\star_2$ and $\star_3$, appearing in (\ref{SWmap}), are defined, respectively, as
\begin{eqnarray}
f(x)\star_2 g(x)&=&[f(x) \stackrel{\star}{,}g(x)]
\nonumber\\
&=&\frac{\sin\frac{\partial_1\theta
\partial_2}{2}}{\frac{\partial_1\theta
\partial_2}{2}}f(x_1)g(x_2)\bigg|_{x_1=x_2=x}\,,
\label{f*2g}
\end{eqnarray}
\begin{eqnarray}
(f(x)g(x)h(x))_{\star_3}&=&\Bigg(\frac{\sin(\frac{\partial_2\theta
\partial_3}{2})\sin(\frac{\partial_1\theta(\partial_2+\partial_3)}{2})}
{\frac{(\partial_1+\partial_2)\theta \partial_3}{2}
\frac{\partial_1\theta(\partial_2+\partial_3)}{2}}
\nonumber\\
&+&\{1\leftrightarrow 2\}\,\Bigg)f(x_1)g(x_2)h(x_3)\bigg|_{x_i=x},
\nonumber\\
\label{fgh*3}
\end{eqnarray}
where $\star$ is associative but noncommutative, while $\star_2$ and $\star_3$
are both commutative but nonassociative. 

\begin{figure}
\begin{center}
\includegraphics[width=8cm,height=3cm]{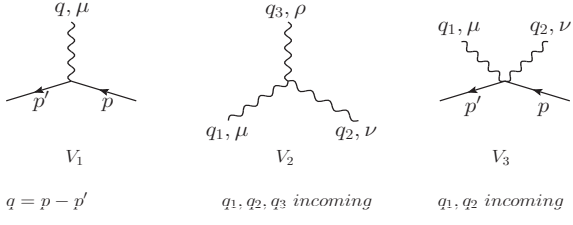}
\end{center}
\caption{Three- and your-field vertices}
\label{fig:Bayzell11Vert}
\end{figure}

The resulting expansion defines
the one-photon-two-fermion, the two-photon-two-fermion and the three-photon vertices,
$\theta$-exactly:
\begin{eqnarray}
S_g&=&\int \;i\partial_\mu a_\nu\star[a^\mu\stackrel{\star}{,}a^\nu]
\nonumber\\
&+&\frac{1}{2}\partial_\mu
\left(\theta^{\rho\sigma}a_\rho\star_2(\partial_\sigma a_{\nu}+f_{\sigma\nu})\right)\star
f^{\mu\nu}.
\label{Sgauge}
\end{eqnarray}
The photon-fermion interaction up to 2-photon-2-fermion terms can derived
by using the first order gauge field and the second order fermion field expansion,
\begin{eqnarray}
S_f&=&\int \;\bar\psi\gamma^\mu[a_\mu\stackrel{\star}{,}\psi]
+(\theta^{ij}\partial_i\bar\psi \star_2
a_j)\gamma^\mu[a_\mu\stackrel{\star}{,}\psi]
\nonumber\\
&+&
i(\theta^{ij}\partial_i\bar\psi
\star_2 a_j)\fmslash\partial\psi-i\bar\psi\star
\fmslash\partial(\theta^{ij}
a_i\star_2\partial_j\psi)
\nonumber\\
&-&
\!\bar\psi\gamma^\mu[a_\mu\!\stackrel{\star}{,}\!\theta^{ij}
a_i\!\star_2\!\partial_j\psi]\!
\nonumber\\
&-&
\!\bar\psi\gamma^\mu
[\frac{1}{2}\theta^{ij}a_i\!\star_2\!(\partial_j
a_\mu\!+\!f_{j\mu})\!\stackrel{\star}{,}\!\psi]\!
\nonumber\\
&-&
\!i(\theta^{ij}\partial_i\bar\psi
\!\star_2\!a_j)\fmslash\partial(\theta^{kl}
a_k\!\star_2\!\partial_l\psi)
\nonumber\\
&+&
\frac{i}{2}\theta^{ij}\theta^{kl}
\big[(a_k\star_2(\partial_l
a_i+f_{li}))\star_2\partial_j\bar\psi
\nonumber\\
&+&
2a_i\star_2(\partial_j(a_k\star_2\partial_l\bar\psi))-
a_i\star_2(\partial_k
a_j\star_2\partial_l\bar\psi)
\nonumber\\
&+&
\big(a_i\partial_k\bar\psi(\partial_j
a_l+f_{jl})-\partial_k\partial_i\bar\psi a_j a_l\big)_{\star_3}\big]
\fmslash\partial\psi
\nonumber\\
&+&
\frac{i}{2}\theta^{ij}\theta^{kl}\bar\psi\fmslash\partial
\big[(a_k\star_2(\partial_l
a_i+f_{li}))\star_2\partial_j\psi
\nonumber\\
&+&
\!2a_i\!\star_2\!
(\partial_j(a_k\!\star_2\!\partial_l\psi))\!-\!a_i\!\star_2\!(\partial_k
a_j\!\star_2\!\partial_l\psi)\!
\nonumber\\
&+&
\big(a_i\partial_k\psi(\partial_j
a_l\!+\!f_{jl})\!-\!\partial_k\partial_i\psi a_j
a_l\big)_{\star_3}\big]\,.
\label{Sfermion}
\end{eqnarray}
Note that actions for gauge and matter fields obtained above,
(\ref{Sgauge}) and (\ref{Sfermion}) respectively, are nonlocal
objects due to the presence of the (generalized) 
star products: $\star$, $\star_2$ and $\star_3$.
Feynman rules from above actions, represented in Fig.\ref{fig:Bayzell11Vert},
are given explicitly in \cite{arXiv:1111.4951}.
\\

\begin{figure}
\begin{center}
\includegraphics[width=8cm,height=4cm]{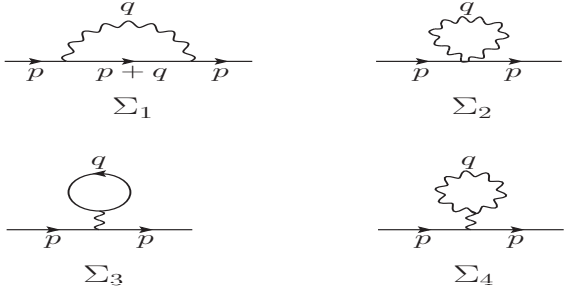}
\end{center}
\caption{One-loop self-energy of a massless neutrino}
\label{Sigma1-loop}
\end{figure}

{\it 3. Neutrino two-point function}\\
As depicted in Fig. \ref{Sigma1-loop}, there are four Feynman diagrams
contributing  to the  $\nu$-self-energy at one-loop.
With the aid of (\ref{Sfermion}), 
we have verified by explicit calculation that the 4-field
tadpole ($\Sigma_2$) does vanish. The 3-fields tadpoles ($\Sigma_3$ 
and $\Sigma_4$) can be ruled out by invoking
the NC charge conjugation symmetry \cite{Aschieri:2002mc}.
Thus only the $\Sigma_1$ diagram needs to be evaluated.   
In spacetime of the dimensionality $D$ we obtain
\begin{eqnarray}
\Sigma_1
&=&\mu^{4-D}\int \frac{d^D q}{(2\pi)^D}
\bigg(\frac{\sin\frac{q\theta p}{2}}{\frac{q\theta
p}{2}}\bigg)^2\frac{1}{q^2}\frac{1}{(p+q)^2}
\nonumber\\
&\cdot & \{(q\theta p)^2(4-D)(\fmslash p+\fmslash q)
\label{Sigma1}\\
&+&(q\theta p)\left[/\!\!\tilde q(2p^2+2p\cdot q)-/\!\!\tilde
p(2q^2+2p\cdot q)\right]
\nonumber\\
&+&\big[\fmslash p(\tilde q^2(p^2+2p\cdot q)-q^2(\tilde p^2+2\tilde
p\cdot\tilde q) )
\nonumber\\
&+&
\fmslash q(\tilde p^2(q^2+2p\cdot q)-p^2(\tilde
q^2+2\tilde p\cdot\tilde q ))\big]\}\,,
\nonumber
\end{eqnarray}
where ${\tilde p}^\mu=(\theta p)^\mu=\theta^{\mu\nu} p_\nu $,
and in addition ${\tilde{\tilde p}}^\mu=(\theta\theta p)^\mu
=\theta^{\mu\nu}\theta_{\nu\rho}p^\rho$.
To perform computations of those integrals using the dimensional
regularization method, we first use
the Feynman parametrization on the quadratic denominators,
then the Heavy Quark Effective theory (HQET)
parametrization \cite{Grozin:2000cm} is used to combine
the quadratic and linear denominators.
In the next stage we use the Schwinger
parametrization to turn the denominators
into Gaussian integrals. Evaluating the relevant integrals for
$D=4-\epsilon$ in the limit $\epsilon\to 0$, we obtain the 
closed form expression for the self-energy as
\begin{eqnarray}
\Sigma_1&=&\gamma_{\mu}
\bigg[p^{\mu}\: A+(\theta{\theta p})^{\mu}\;\frac{p^2}{(\theta p)^2}\;B\bigg]\,,
\label{sigma1AB}\\
A&=& \frac{-1}{(4\pi)^2}
\bigg[p^2\;\bigg(\frac{\tr\theta\theta}{(\theta p)^2}
+2\frac{(\theta\theta p)^2}{(\theta p)^4}\bigg) A_1 
\nonumber\\
&+&
 \bigg(1+p^2\;\bigg(\frac{\tr\theta\theta}{(\theta p)^2}
+\frac{(\theta\theta p)^2}{(\theta p)^4}\bigg)\bigg)A_2 \Bigg]\,,
\label{A}\\
A_1&=&\frac{2}{\epsilon}+\ln(\mu^2(\theta p)^2)
+ \ln({\pi e^{\gamma_{\rm E}}})
\label{A1}\\
&+&
\sum\limits_{k=1}^\infty \frac{\left(p^2(\theta p)^2/4\right)^k}{\Gamma(2k+2)}
\left(\ln\frac{p^2(\theta p)^2}{4} + 2\psi_0(2k+2)\right) \,,
\nonumber\\
A_2&=&-\frac{(4\pi)^2}{2} B = -2
\nonumber\\
&+& \sum\limits_{k=0}^\infty
\frac{\left(p^2(\theta p)^2/4\right)^{k+1}}{(2k+1)(2k+3)\Gamma(2k+2)}
\Bigg(\ln\frac{p^2(\theta p)^2}{4} 
\nonumber\\
&-&
2\psi_0(2k+2)
- \frac{8(k+1)}{(2k+1)(2k+3)} \Bigg),
\label{A2}
\end{eqnarray}
with  $\gamma_{\rm E}\simeq0.577216$ being Euler's constant.

The $1/\epsilon$ UV divergence
could in principle be removed by a properly chosen counterterm.
However (as already mentioned) due to the specific momentum-dependent 
coefficient in front of it, a nonlocal form for it is required. 

It is important to stress here that amongst other
terms contained in both coefficients $A_1$ and $A_2$, there are structures
proportional to
\begin{equation}
\left({p^2(\theta p)^2}\right)^{n+1}(\ln{(p^2(\theta p)^2)})^m,\;\;
{\forall n} \;\; {\rm and} \;\;m=0,1.
\label{lnUV/IR'}
\end{equation}
The numerical factors in front of the above structures are rapidly-decaying,
thus series are always convergent for finite argument, 
as we demonstrate in \cite{arXiv:1111.4951}.\\

{\it 4. UV/IR mixing}\\
Turning to the UV/IR mixing problem, we recognize
a soft UV/IR mixing term represented by a logarithm,
\begin{equation}
\Sigma_{\rm UV/IR}=-{\fmslash p}\,p^2 \,
\bigg(\frac{\tr\theta\theta}{(\theta p)^2}
+2\frac{(\theta\theta p)^2}{(\theta p)^4}\bigg)\cdot
\frac{2}{(4\pi)^2}\ln{|\mu(\theta p)|}.
\label{lnUV/IR}
\end{equation}
Instead of dealing with nonlocal counterterms, we take a different route
here to  cope with various divergences besetting (\ref{sigma1AB}). Since $\theta^{0i}
\neq 0$ makes a NC theory nonunitary \cite{Gomis:2000zz},
we can, without loss of generality,
chose $\theta$ to lie in the (1, 2) plane
\begin{equation}
\theta^{\mu\nu}=\frac{1}{\Lambda_{\rm NC}^2}
\begin{pmatrix}
0&0&0&0\\
0&0&1&0\\
0&-1&0&0\\
0&0&0&0
\end{pmatrix}.
\label{degen}
\end{equation}
Automatically, this produces
\begin{equation}
\frac{\tr\theta\theta}{(\theta p)^2}
+2\frac{(\theta\theta p)^2}{(\theta p)^4}=0,\:\forall p.
\label{degen1}
\end{equation}
With (\ref{degen1}), $\Sigma_1$,
in terms of Euclidean momenta, receives the following form:
\begin{equation}
\Sigma_1=\frac{-1}{(4\pi)^2}\gamma_\mu\left[p^\mu
\bigg(1 + \frac{\tr\theta\theta}{2}\frac{p^2}{(\theta p)^2}\bigg)
-2(\theta\theta p)^\mu\frac{p^2}{(\theta p)^2} \right] A_2.
\label{Sigmabuble}
\end{equation}
By inspecting (\ref{A2}) one can be easily convinced that $A_2$ is 
free from the $1/\epsilon$ divergence and the UV/IR mixing term, being also 
well-behaved in the infrared, in the $\theta \rightarrow 0$ 
as well as $\theta p \rightarrow 0$
limit. We see, however, that the two terms in (\ref{Sigmabuble}), 
one being proportional to
$\fmslash{p}$ and the other proportional to $\fmslash{\tilde{\tilde p}}$,
are still ill-behaved in the $\theta p \rightarrow 0$
limit. If, for the choice (\ref{degen}), $P$ denotes the momentum in the (1, 2)
plane, then $\theta p = \theta P$. For instance, a particle moving
inside the NC plane with
momentum $P$ along the one axis, has a spatial extension of size $|\theta P|$
along the other. For the choice (\ref{degen}), $\theta p \rightarrow 0$ corresponds to
a zero momentum projection onto the (1, 2) plane. Thus, albeit in our approach the
commutative limit ($\theta \rightarrow 0$) is smooth at the quantum level,
the limit when an extended object (arising due to the fuzziness of space)
shrinks to zero, is not. We could surely claim that in our approach the
UV/IR mixing problem is considerably softened; on the other hand, we have
witnessed how the problem strikes back in an unexpected way. This is, at the
same time, the first example where this two limits are not degenerate.\\

{\it 5. Neutrino dispersion relations}\\
In order to probe physical consequence of
the 1-loop quantum correction, with $\Sigma_{1-loop_{M}}$ 
from Eq. (3.25) in \cite{arXiv:1111.4951}, 
we consider the modified propagator
\begin{equation}
\frac{1}{\fmslash\Sigma}=\frac{1}{\fmslash p-\Sigma_{1-loop_{M}}}=
\frac{\fmslash\Sigma}{\Sigma^2}\,.
\label{disp}
\end{equation}
We further choose the NC parameter to be (\ref{degen})
so that the denominator is finite and can be expressed explicitly:
\begin{equation}
\hspace{-.05cm}\Sigma^2=p^2\left[\hat A_2^2\left(\frac{p^4}{p^4_r}
+2\frac{p^2}{p_r^2}+5\right)-\hat A_2\left(6+2\frac{p^2}{p_r^2}\right)+1\right],\,(23)
\nonumber
\end{equation}
where $p_r$ represents $r$-component of the momentum $p$
in a cylindrical spatial coordinate system and $\hat A_2= e^2A_2/(4\pi)^2=-B/2$.

From above one see that $p^2=0$ defines one set of the dispersion relation,
corresponding to the dispersion for the massless neutrino mode,
however the denominator $\Sigma^2$ has one more coefficient $\Sigma'$
which could also induce certain zero-points. Since the $\hat A_2$ is a
function of a single variable $p^2p_r^2$, with $
p^2=p^2_0-p^2_1-p^2_2-p^2_3\,\,\rm and\,\,p^2_r=p^2_1+p^2_2$,
the condition $\Sigma'=0$  can be expressed as a simple algebraic equation
\begin{equation}
\hat A_2^2z^2-2\left(A_2-\hat A_2^2\right)z
+\left(1-6\hat A_2+5\hat A_2^2\right)=0\,,
\label{A2z2}
\end{equation}
of new variables $z:=p^2/p_r^2$, in which the coefficients
are all functions of $y:=p^2p^2_r/\Lambda^4_{\rm NC}$.

The two formal solutions of the equation (\ref{A2z2})
\begin{equation}
z=\frac{1}{\hat A_2}\left[\left(1-\hat A_2\right)\pm\,2\left(\hat A_2
-\hat A_2^2\right)^{\frac{1}{2}}\right]\,,
\label{zsolutions}
\end{equation}
are birefringent.
The behavior of solutions (\ref{zsolutions}), is next analyzed at
two limits $y\to \,0$, and $y\to\,\infty$.\\

\noindent
{\it The low-energy regime: $p^2p^2_r \ll\Lambda^4_{\rm NC}$}\\
\noindent
For $y \ll 1$ we set $\hat A_2$ to its zeroth order value $ e^2/8\pi^2$, 
\begin{eqnarray}
p^2&\sim& \left(\left(\frac{8\pi^2}{e^2}-1\right)\pm 2
\left(\frac{8\pi^2}{e^2}-1\right)^{\frac{1}{2}}\right)\cdot\,p_r^2
\nonumber\\
&\simeq&\left(859\pm 59\right)\cdot\,p_r^2\,,
\label{859pm59}
\end{eqnarray}
obtaining two (approximate) zero points. From the definition of $p^2$
and $p_r^2$ we see that both solutions are real and positive.
Taking into account the higher order (in y) correction
these poles will locate nearby the real axis of
the complex $p_0$ plane thus correspond to some metastable modes with
the above defined dispersion relations. As we can see, 
the modified dispersion relation \eqref{859pm59} does not depend on 
the noncommutative scale, therefore it introduces a discontinuity in 
the $\Lambda_{\rm NC}\to\infty$ limit, 
which is not unfamiliar in noncommutative theories.\\

\noindent
{\it  The high-energy regime: $p^2p^2_r \gg \Lambda^4_{\rm NC}$}\\
At $y\gg 1$ we analyze the asymptotic behavior of
\begin{equation}
A_2\sim \frac{i\pi^2}{8} {\sqrt y}\left(1-\frac{16i}{\pi y}
e^{-\frac{i}{2}{\sqrt y}}\right)+\mathcal O\left(y^{-1}\right),
\end{equation}
from \cite{arXiv:1111.4951}, therefore \eqref{zsolutions} can be reduced to
\begin{equation}
z\sim -1\pm 2i \;\; \rightarrow \;\;p^2_0\sim p^2_3\pm 2i p^2_r.
\end{equation}
We thus reach two unstable deformed modes besides the usual mode $p^2=0$ in the high energy regime. Here again the leading order deformed dispersion relation does not depend on the noncommutative scale $\Lambda_{\rm NC}$.
\\

{\it 6. Neutrino two-point function from alternative action}
\\
Using the Feynman rule of the alternative action (2.15) from Ref \cite{arXiv:1111.4951},
we find the following contribution
to the neutrino self-energy from diagram $\Sigma_1$
\begin{equation}
\Sigma_{1_{alt}}=\frac{\fmslash p}{(4\pi)^2} \bigg[\frac{8}{3} \frac{1}{{(\theta p)^2}}
\bigg(\frac{\tr\theta\theta}{(\theta p)^2}
+4\frac{(\theta\theta p)^2}{(\theta p)^4}\bigg)\bigg]\,.
\label{A9}
\end{equation}
The detailed computation is presented in  Ref. \cite{arXiv:1111.4951} (Appendix B). 
Using the same technique and the choice (\ref{degen}) 
we have found modified propagator of the following form
\begin{equation}
\frac{i}{\fmslash \Sigma_{1_{alt}}}= 
\frac{i}{\fmslash p} \frac{p_r^4}{p_r^4+\frac{\Lambda^4_{\rm NC}}{3\pi^2}}
=\Pi  \frac{i\fmslash p}{p^2}.
\label{A10}
\end{equation}
Since the only pole for $p_0$ comes from $p^2=0$, there is no modified 
dispersion relation. However, we can see that $\Pi =0$ when $\vert p_r \vert= 0$,
therefore there could be directional dependent effect on the propagation. 
One may try a source function $f(x_\mu)$, then 
(neglecting some details about time ordering) for a wave function $\phi(x)$ 
which propagates according to the scalar part of the modified propagator \eqref{A10} one could have
\begin{equation}
\begin{split}
\phi(x_\mu)\sim&\int d^4x'\int d^4p\frac{i\Pi}{p^2}e^{ip(x-x')}f(x')
\\
\sim&\int d^4p \frac{i\Pi}{p^2}e^{ipx}\tilde f(p).
\end{split}
\end{equation}
The direction dependent factor $\Pi$ could then take zero value for $|p_r|=0$ modes, thus give suppressive effect.

We have to notice that there are no hard
$1/\epsilon$ UV divergent and no logarithmic UV/IR mixing terms,
and the finite terms like in $A_1$ and $A_2$ are also absent.
Thus the subgraph $\Sigma_1$ does not require any counter-term.
However, the result (\ref{A9}), does express powerful UV/IR mixing effect, 
that is in terms of scales terms, the $\Sigma_{1_{alt}}$ experience
the forth-power of the {\it NC-scale/momentum-scale} ratios
$\sim |p|^{-2}|\theta p|^{-2}$ in (\ref{A9}), i.e. we are dealing with
the $\Sigma_{1_{alt}}\sim{\fmslash p}\left({\Lambda_{\rm NC}}/{p}\right)^4$
within the ultraviolet and infrared limits for $\Lambda_{\rm NC}$ and $p$, 
respectively.\\

{\it 7. Discussion and conclusions}\\
If new physics were  capable to push the  neutrino-nucleon inelastic cross
section three orders of magnitude beyond the standard-model (SM) 
prediction, then
ultra-high energy (UHE) neutrinos would have already been observed at 
neutrino observatories. 
We use such a constraint to reveal information on the scale
of noncommutativity $\Lambda_{\rm NC}$ in 
noncommutative gauge field theories (NCGFT) where
$\nu$'s possess a tree-level coupling to photons in a
generation-independent manner. In the energy range of interest ($10^{7}$ to
${10^8}$ TeV for example)  where there is always energy of 
the system $(E)$ larger than the NC scale $(E/\Lambda_{\rm NC}>1)$,
the perturbative expansion in terms of $\Lambda_{\rm NC}$ retains 
no longer its meaningful character, thus it is
forcing us to resort to those NC field-theoretical frameworks involving the 
full $\theta$-resumation. Our
numerical analysis of the contribution to the process coming from the photon
exchange, pins impeccably down a lower bound on $\Lambda_{\rm NC}$ 
to be as high as around up to ${\cal O}(10^3)$ TeV,
depending on the cosmogenic $\nu$-flux. 

We should stress that from the same actions (\ref{Sgauge}, \ref{Sfermion}), 
but for three different cosmological laboratories, 
that is from UHE cosmic ray neutrino scatterings on nuclei \cite{Horvat:2010sr}, 
from the BBN nucleosynthesis and from the reheating phase after inflation \cite{Horvat:2009cm},
we obtain very similar, a quite strong bounds on the NC scale, 
of the order of $10^3$ TeV. Note in particular that all results depicted 
in Figs.\ref{fig:ncSM-CrossSections}-\ref{fig:LambdaVsT}, 
show closed-convergent forms.

Next we discuss $\theta$-exact computation of the
one-loop quantum correction to the $\nu$-propagator.
We in particular evaluate  the neutrino two-point function, 
and demonstrate how quantum effects in the $\theta$-exact 
SW map approach to NCGFT's, together with a combination of Schwinger,
Feynman, and  HQET parameterization, reveal a much richer structure
yielding the one-loop quantum correction in a closed form.
 
The closed form general expression for the neutrino self-energy
(\ref{sigma1AB}) contains in (\ref{A1}) both a hard $1/\epsilon$
UV term and  the celebrated UV/IR mixing term
with a logarithmic infrared singularity $\ln\sqrt{(\theta p)^2}$.
Results shows complete decoupling of the UV divergent term 
from softened UV/IR mixing term and from the finite terms as well. 
Our deformed dispersion relations at both the low and high energies 
and at the leading order
do not depend on the noncommutative scale $\Lambda_{\rm NC}$.

The low energy dispersion relation \eqref{859pm59} is, 
in principle, capable of generating a direction dependent superluminal velocity, 
this can be seen clearly from the maximal attainable velocity of the neutrinos
\begin{equation}
\frac{{\rm v}_{max}}{c}=\frac{dE}{d|\vec p|}\sim \sqrt{
1+\left(859\pm 59\right)
\sin^2\vartheta}\,,
\label{varepsilon}
\end{equation}
where $\vartheta$ is the angle with respect to the direction perpendicular 
to the NC plane. This gives one more example how such spontaneous breaking 
of Lorentz symmetry (via the $\theta$-background) could affect 
the particle kinematics through quantum corrections 
(even without divergent behavior like UV/IR mixing). 
On the other hand one can also see that the magnitude of superluminosity 
is in general very large in our model, thus seems contradicting various observations.
The currently largest superluminal velocity for neutrinos, 
$({\rm v} -c)/c = (2.37 \pm 0.32 ({\rm stat.}){+0.34 \atop -0.24} ({\rm sys.})) 
\times 10^{-5}$ reported by the OPERA collaboration \cite{:2011zb}, 
has been found to be suffering from multiple technical difficulties lately 
\cite{OPERAnews}. Other experiments suggests much smaller values 
\cite{Hirata:1987hu,Bionta:1987qt,Longo:1987ub}. 
The authors consider here, on the other hand, 
that the large superluminal velocity issue may be reduced/removed by 
taking into account several further considerations:\\
(a) First, the reason to select a constant nonzero $\theta$ 
background in this paper is computational simplicity. 
The results will, however, still hold for a NC background 
that is varying sufficiently slowly with respect to the scale of
noncommutativity. There is no physics reason to expect 
$\theta$ to be a globally constant background {\it ether}.  
In fact, if the $\theta$ background is only nonzero in 
tiny regions (NC bubbles) the effects of the modified dispersion
relation will be suppressed macroscopically.
Certainly a better understanding of possible sources of NC is needed.\\
(b) Second, in our computation we considered only the purely noncommutative 
neutrino-photon coupling, it has been pointed out that modified 
neutrino dispersion relation could open decay channels within 
the commutative standard model framework \cite{Cohen:2011hx}. 
In our case this would further provide decay channel(s) which can 
bring superluminal neutrinos to normal ones.\\
(c) Third, as we have stated in the introduction, 
model 1 is not the only allowed deformed model with noncommutative 
neutrino-photon coupling. And as we have shown for our model 2, 
there could be no modified dispersion relation(s) for deformation(s) 
other than 1, therefore it is reasonable to conjecture 
that Seiberg-Witten map freedom may also serve as one possible 
remedy to this issue.

The essential difference of our results as compared to \cite{Hayakawa:1999yt}
is that in our case both terms are proportional to the spacetime
noncommutativity dependent $\theta$-ratio factor in (\ref{degen1}),
which arise from the natural non-locality of our actions and does not depend on
the noncommutative scale, but only on the scale-independent structure
of the noncommutative $\theta$-ratios.
Besides the divergent terms, a new spinor structure $(\theta\theta p)$
with finite coefficients emerges in our computation, see (\ref{sigma1AB})-(\ref{A2}).
All these structures are proportional to $p^2$, therefore if appropriate
renormalization conditions are imposed, the commutative dispersion relation
$p^2=0$ can still hold, as a part of the full set of solutions obtained in (23).

Finally, we mention that our
approach to UV/IR mixing should not be confused with the one  based on
a theory with UV completion ($\Lambda_{\rm UV} < \infty$), where a theory
becomes an effective QFT, and the UV/IR mixing manifests itself via a
specific relationship between the UV and the IR cutoffs
\cite{AlvarezGaume:2003mb,Abel:2006wj}.

The properties summarized above are previously unknown 
features of $\theta$-exact NC gauge field theory. 
They appear in the model with the action presented in section~2.
The alternative action, and the corresponding $\nu$-self-energy (\ref{A9}),
has less striking features, but it does have it's own advantages due to 
the absence of a hard UV divergences, and the
absence of complicated finite terms. The structure in (\ref{A9}) is
different (it is {\it NC-scale/energy} dependent) with respect to
the NC scale-independent structure from (\ref{degen1}),
as well as to the structure arising from fermion self-energy computation in
the case of $\star$-product only unexpanded theories 
\cite{Hayakawa:1999yt,Brandt:2001ud}.
However, (\ref{A9}) does posses powerful UV/IR mixing effect.
This is fortunate with regard to the use of  low-energy NCQFT
as an important window  to quantum gravity \cite{Szabo:2009tn} 
and holography \cite{Horvat:2010km}.
\section*{Acknowledgment}
Work supported by the Croatian Ministry of Science, Education and
Sport project 098-0982930-2900. I would like to thank Jiangyang You for many
valuable comments/remarks. Author  would also like to acknowledge support of 
Alexander von Humboldt Foundation (KRO 1028995), as well as to thanks W. Hollik and the Max-Planck-Institute for Physics, Munich, Germany,  for hospitality.\\

*Based on the plenary session talk given at the XXIX Max Born Symposium: Wroclaw-Poland 2011, "Super, Quantum \& Twistors II", Wroclaw, 28-30 June 2011.

\end{document}